# Epitaxial $Bi_2FeCrO_6$ Multiferroic thin films

Riad Nechache[1], Catalin Harnagea[1], Louis-Philippe Carignan[2], David Ménard[2], and Alain Pignolet[1]

[1] INRS - Énergie, Matériaux et Télécommunications,

1650, boulevard Lionel-Boulet,

Varennes (Québec), J3X 1S2 (Canada).

E-mail : pignolet@emt.inrs.ca

[2] École Polytechnique de Montréal,

Département de Génie Physique,

P.O. Box 6079, Station. Centre-ville,

Montréal, (Québec), H3C 6A7 (Canada).

E-mail : David.Menard@polymtl.ca

# ABSTRACT

We present here experimental results obtained on $Bi_2FeCrO_6$ (BFCO) epitaxial films deposited by laser ablation directly on $SrTiO_3$ substrates. It has been theoretically predicted, by Baettig and Spaldin, using first-principles density functional theory that BFCO is ferrimagnetic (with a magnetic moment of $2\mu_B$ per formula unit) and ferroelectric (with a polarization of ~80 µC/cm$^2$ at 0K). The crystal structure has been investigated using X-ray diffraction which shows that the films are epitaxial with a high crystallinity and have a degree of orientation depending of the deposition conditions and that is determined by the substrate crystal structure. Chemical analysis carried out by X-ray Microanalysis and X-ray Photoelectron Spectroscopy (XPS) indicates the correct cationic stoichiometry in the BFCO layer, namely (Bi:Fe:Cr = 2:1:1). XPS depth profiling revealed that the oxidation state of Fe and Cr ions in the film remains 3+ throughout the film thickness and that both Fe and Cr ions are homogeneously distributed throughout the depth. Cross-section high-resolution transmission electron microscopy images together with selected area electron diffraction confirm the crystalline quality of the epitaxial BFCO films with no identifiable foreign phase or inclusion.

The multiferroic character of BFCO is proven by ferroelectric and magnetic measurements showing that the films exhibit ferroelectric and magnetic hysteresis at room temperature. In addition, local piezoelectric measurements carried out using piezoresponse force microscopy (PFM) show the presence of ferroelectric domains and their switching at the sub-micron scale.

## 1. Introduction

Multiferroics are materials possessing at least two order parameters (e.g. spontaneous magnetization, spontaneous polarization or spontaneous strain) co-existing simultaneously and possibly exhibiting a coupling among them. Magnetic ferroelectrics are rare in nature since transition metal ions with unpaired magnetically active d-electrons often tend to reduce the off-center distortion necessary for ferroelectricity to exist.[1] Different families of multiferroic systems have been identified among which $RMnO_3$ (R = Dy, Tb, Ho, Y, Lu, etc.) with hexagonal structure, $RMn_2O_5$ (e.g. R = Nd, Sm) with orthorhombic structure, $BiMO_3$–type (e.g. M = Mn, Fe) with perovskite or pseudo-perovskite structure (among which the double-perovskite $Pb_2(CoW)O_6$), and $M_3B_7O_{13}X$ (with e.g. M = Cr, Mn, Fe, Ni … and X = Cl, Br, I) having a boracite structure. Such materials may present magnetoelectric coupling and therefore, ideally, magnetization can be controlled by an electric field and vice versa. Recently, there has been a surge in research interest for multiferroic systems due to the observation of good multiferroic properties and magnetoelectric coupling in thin films of some of these materials. This enables potential applications in information storage and spintronics such as magneto-electric sensors[2], magneto-capacitive devices, and electrically driven magnetic data strorage and recording devices. However, the known 'ferromagnetic ferroelectrics', like $RMnO_3$ (space group P63cm), $RMn_2O_5$ (space group Pbam), $BiMnO_3$ (space group C2) or $BiFeO_3$ (space group R3c) tend to have low magnetic Curie temperatures[3], have often very weak ferromagnetic properties,[4] or are not strong enough insulators at room temperature to reliably measure the ferroelectric properties, and thus to make use of the spontaneous polarization.[5]

Using first-principles density functional theory, Spaldin et al.[6] described the design of a new multiferroic material, namely $Bi_2FeCrO_6$ (BFCO). They predicted that this material would be ferrimagnetic at 0 K with a magnetic moment of $2\mu_B$ per formula unit and would be ferroelectric with a polarization of 80 µC/cm$^2$. They also predicted that BFCO would have a rhombohedral structure very similar to that of BFO, except that every second iron cation is replaced by a chromium cation in the (111) direction, which reduces the symmetry of the space group to R3.

Motivated by those ab-initio calculations, we recently reported good multiferroic properties at room temperature of epitaxial PLD-grown films of BFCO, deposited either on epitaxial SrRuO$_3$ (SRO) films on (100)-oriented SrTiO$_3$ (STO) single crystalline substrates.[7]

We report here the successful fabrication of BFCO epitaxial films directly on STO by pulsed laser deposition (PLD) and we present their structural, electric and magnetic properties characterization as well as an analysis of their chemical composition. These results are compared to the properties predicted by *ab-initio* calculations and the differences are analyzed and discussed. Due to the similarity of the crystal structure of BFCO with that of BiFeO$_3$ (BFO), the results of our study are systematically compared with properties of good quality BFO thin films that are used as benchmark.

## 2. Experimental

### 2.1 Pulsed Laser Deposition

The BFCO films were prepared by pulsed laser deposition (PLD) using a pulsed KrF excimer laser (λ= 248 nm) with a fluence of 2.0 J cm$^{-2}$ and a pulse repetition rate of 8Hz. A dense ceramic target composed of a mixture of 50 mol.% BiFeO$_3$ and 50 mol.% BiCrO$_3$ was utilized. The BCFO films were deposited on (100)-oriented SrTiO$_3$ single crystalline substrates doped with 0.5 wt% Nb (STO:Nb) (from CrysTec GmbH). The deposition conditions used for the deposition of BFCO have been derived from the deposition conditions that were earlier optimized for the deposition of BFO. In an earlier study, we deposited BFO films on STO:Nb with different thicknesses and optimized the conditions found in the literature to obtain the best BFO structural quality.[21,22] The optimum deposition conditions for our BFO films were then used as guidelines and were further refined and optimized for BFCO deposition. The substrate temperature during the films deposition was set to 700°C and the BFCO films were deposited at a pressure of 10 mTorr of oxygen resulting in a growth rate of ~10 nm/min. To eliminate possible oxygen vacancies, the films were slowly cooled from 700°C down to 400°C at a cooling rate of 8°C/min in an atmosphere of oxygen and maintained at 400°C for one hour, prior cooling down to room temperature. The films thicknesses were measured using a variable angle spectroscopic ellipsometer (J. A. Woollam Co) with a photon energy between 0.1 and 3 eV. The film thicknesses were simultaneously measured by X-ray reflectometry (XRR) and the two measurements were compared (for the films having a thickness less than 150 nm, the maximum thickness measured by our XRR setup).

### 2.2 Chemical and Structural Analysis

Rutherford Backscattering Spectrometry (RBS) (A350 keV Van de Graff accelerator) and energy dispersive x-ray spectroscopy (EDX) (Oxford Instruments) installed inside a Scanning Electron Microscope (SEM) (JEOL JSM-6300F) were used to analyze the BFCO film stoichiometry. Both techniques indicate that the sample is chemically homogeneous with a correct cationic stoichiometry Bi:Fe:Cr = 2:1:1.

The oxidation states of the cations as well as the composition were investigated across the whole layer's depth for a 80 nm thick film using X-ray photoelectron spectroscopy (XPS) (ESCALAB 220i-XL system). Depth profiling was performed with intermittent $Ar^+$ ion sputtering. The sputtering was done with a 3-KeV $Ar^+$ ion beam (current 0.1-0.2 µA) in an Ar pressure of 2-3 $\times 10^{-8}$ Torr. The ion beam was scanned over an area of 2 × 2 $mm^2$. Under the conditions used, the average sputtering rate measured was about 0.033 nm/sec. The x-ray photoelectron Bi4f, Fe2p, Cr2p and O1s core level spectra of the BFCO successive surface layers were taken with hemispherical electron energy analyzer using a Mg K$_\alpha$ X-rays twin source (1253.6 eV). The pressure inside the analysis chamber during XPS measurements was maintained below 3 X10$^{-9}$ Torr. The atomic concentration was determined by using sensitivity factors supplied by the instrument manufacturer. During XPS measurement, the binding energy scale was corrected and calibrated using the Bi4$f_{7/2}$ peak position (158.8 eV). For comparison, we performed the same measurement on a 90 nm thick BFO film.

Structural characterization of BFCO films was performed using x-ray diffraction (XRD) (PANalytical X'Pert MRD 4-circle diffractometer). High-resolution transmission electron microscopy (HRTEM) (JEOL JEM-2100F) and electron diffraction analysis were also employed to characterize the structural properties of the BFCO layer.

### 2.3 Dielectric properties

To test the local ferroelectric properties of the BFCO films, piezoresponse force microscopy (PFM) [8,9,10] was used to image and manipulate the ferroelectric polarization of these films. Here we used a DI-Enviroscope AFM (Veeco) equipped with a NSC36a (Micromasch) cantilever and tips coated with Co/Cr. We applied an ac voltage of 0.5V at 26 kHz between the conductive tip and the STO:Nb conducting substrate located beneath the BFCO layer and we detected surface induced piezoelectric vibrations using a Lock-in Amplifier from Signal Recovery (model 7265).

### 2.4 Magnetic properties

The magnetic hysteresis (M-H) loops were measured at room temperature using a vibrating-sample magnetometer (VSM) (ADE Technologies). A maximum saturating field of 10 kOe was applied parallel to film plane, then decreased down to -10 kOe by step of 500 Oe, and back up to 10 kOe. The field steps were decreased down to 50 Oe around the zero field range for better resolution. An average factor of 20 measurements per field points provided an absolute sensitivity of about $10^{-6}$ emu, The observed hysteresis clearly indicated the presence of an ordered magnetic phase. The magnetic responses of the sample holding rods and of bare substrates (without BFCO) were also measured in order to confirm that the ferro- or ferrimagnetic signal was originating from the BFCO films.

## 3. Results and Discussion

### 3.1 Chemical Analysis

Figure 1A shows the typical XPS core level spectra of Fe2$p$ and Cr2$p$ lines. The Cr2$p$ XPS spectra reveal the usual 2$p_{3/2}$ and 2$p_{1/2}$ doublets arising from spin-orbit splitting. The binding energies for Cr2$p_{3/2}$ and Cr2$p_{1/2}$ peaks are about 576.9 and 586.7± 0.2 eV, respectively. For comparison, the Cr2$p_{3/2}$ binding energies have been reported to be near 576.5 eV for $Cr^{3+}$ [11] in LaCrO$_3$ (LCO), 574.2 eV for metallic Cr, as well as 576.0 eV for $Cr^{2+}$ and 577.0 eV for $Cr^{3+}$ in Cr$_2$O$_3$(111)/Cr(111) films[12]. It can be seen that the Cr2$p_{3/2}$ peak in the XPS spectra of BFCO is very close to that of Cr in LaCrO$_3$, implying that the oxidation state of the Cr ion on BFCO is $Cr^{3+}$. The Fe2p peak in the BFCO spectra is characterized by a Fe2$p_{3/2}$ peak with maximum situated at 711.0 ± 0.1 eV which agrees well with the value of binding energies for Fe reported for BFO.[13] The broad Fe2p peaks are attributed to $Fe^{3+}$ oxidation states,[14] and no satellite peak is observed between 715 and 720 eV, which excludes the possible presence of γ-Fe$_2$O$_3$ (maghaemite) or of α-Fe$_2$O$_3$ (haematite) phases in our films, where such satellites are generally observed.[15] We can notice also that the broad Fe2p peaks for BFCO is different from the mixed Fe valence state ($Fe^{3+}$ - $Fe^{2+}$) in Fe$_3$O$_4$.[15]

The results clearly show that the depth distribution of Fe and Cr is homogeneous within the whole thickness of the BFCO layer. Using the area of the Fe 2p$_{3/2}$ and Cr2p$_{3/2}$ peaks in the obtained photoelectron spectra, the atomic concentration for both Fe and Cr were determined quantitatively. We found that both concentrations are nearly constant across the film thickness and are approximately 9.69 at% for Fe and 9.67 at% for Cr, close to the nominal 10 at% expected for both (within an experimental error of 3 at%).

Figure 1B shows the O1s core level XPS spectra obtained near the surface (at the depth of ~3-5 nm, so that the surface contamination carbon layer was removed) of the both BFCO and BFO films. Here the charge-shifted spectra were corrected using the adventitous C1s photoelectron signal at 285 eV. The O1s line is composed of two peaks; the binding energies near the surface are 530.2 eV and 532.8 eV for BFCO, and 530.9 eV and 532.8 eV for BFO. This suggests that the oxygen in the near-surface region of the films exists mainly within the BFCO (or BFO) structure with a binding energy of 530.2 eV (respectively 530.9 eV for BFO), and as surface adsorbed oxygen (at 532.8 eV). [24]

For BFCO, The binding energy of the O1s line near the surface region appears to be lower than that of the BFO film by 0.5 eV (with a measurement accuracy of 0.1 eV), indicating a change in the local environment of the oxygen atoms in BFCO with respect to BFO. This shift reveals a higher cation-oxygen bond polarizability for BFCO compared to BFO, attributed to a lower electronegativity for Cr than for Fe. We also note that the full width at half maximum (FWHM) of the O1s peak in BFCO is slightly larger than that of BFO, which could denote the presence of some atomic disorder in the BFCO layer (for instance caused by disorder in the Fe-Cr sequence along <111> or by a slight deviation of the nominal Fe/Cr atomic ratio), possibly due to the surface proximity.

### 3.2 Structural Analysis

Figure 2A present the θ/2θ diffraction pattern of a film grown directly on STO: Nb 100 substrates, and exhibits only the (00$l$) (with $l$ = 1, 2, 3) reflections of BFCO and STO, indicating that the 285 nm thick BFCO film is highly (001)-oriented. We did not observe any reflections that would be indicative of second phases, in particular iron oxides, in agreement with the previous XPS analysis. The degree of in-plane orientation was assessed examining XRD Φ-scan spectra. As presented in Figure 2B, the four peaks for (103) reflections of BFCO film are 90° apart each other. In addition, the peaks for (103) reflections of BFCO layer occur at the same azimuthal Φ angles as those for STO (103) reflections. This clearly indicates the presence of fourfold symmetry for BFCO film and a "cube-on-cube" epitaxial growth on the STO (100) substrate. The out of plane lattice parameter was calculated to be 3.94 Å for a 285 nm thick BFCO film.

The selected area electron diffraction pattern (SAED) (Fig. 3A), obtained from a cross section of the (001)-oriented sample along the [011] direction confirms the good single crystalline quality of the BFCO layer. Indexing of this SAED pattern with pseudocubic indices yields an in-plane lattice parameter very close to that of STO substrate (~3.91 Å) and an out-of-plane parameter of 3.95 Å. The same out-of-plane lattice parameter is given by high resolution TEM (HRTEM) image in figure 3B showing (001) atomic planes.

### 3.3 Multiferroicity in BFCO film

To prove the multiferroic character of our films we performed measurements of both their ferroelectric and magnetic properties. Macroscopic ferroelectric hysteresis loops were similar to those already reported for our BFCO films on SRO buffer layers.[7] They show a relatively low polarization due to a large leakage current. Here we will show only local electromechanical results and macroscopic magnetic measurements.

Ferroelectricity of our films at the grain level is demonstrated by local electromechanical measurements using PFM, as shown in Figure 4. The grain encircled in Fig. 4A showed an initial polarization oriented downward (top to bottom, black in Fig.4B). After switching the polarization several times by taking multiple hysteresis loops (Fig.4D) the grain was imaged again and showed opposite contrast proving that polarization has indeed been switched (Fig.4C). The hysteresis loops in Figure 4D reveal that a single grain of 285nm thick BFCO (about 500 nm in lateral size) switch and that the reversed polarization is stable (at the time scale of the experiment, 3 days). The in-field hysteresis loop shows the signal *while* the electric DC-bias voltage is applied, therefore including an electrostatic contribution arising from the interaction between the cantilever and the bottom electrode. [8] In contrast, the remanent loop records the signal *after* a DC-bias pulse of a given voltage has been *applied and switched off*, therefore being insensitive to interactions of electrostatic nature. The strength of the PFM signal recorded is comparable to that obtained from BFO films. The surface morphology of the as-deposited BFCO film was examined using atomic force microscopy (AFM). The film had small grain size, typically less than 250 nm. A root-mean-square (rms) roughness of about 3 nm for a 10 × 10 $\mu m^2$ surface of film has been measured.

We now turn our attention to the magnetic properties as well as to the effect of the layer thickness on the magnetic response of BFCO epitaxial thin films fabricated directly on STO:Nb (100) substrates. As shown in the inset of Figure 5, epitaxial films of two different thicknesses have an hysteresis at room temperature, indicating the presence of an ordered magnetic phase. The saturation magnetization ($M_s$) of the 285 nm thick BFCO film was found to be ~ 20 emu/cc, which is very close to the value obtained for a 300 nm thick BFCO deposited on heteroepitaxial SRO (60 nm) / STO (100).[7] A higher saturation magnetization of 40 emu/cc was observed in the 350 nm thick film. For BFO films, an increase in macroscopic magnetization with decreasing thickness was attributed either to the formation of $Fe^{2+}$ ions,[16,17,23] to a homogenization of the magnetic spins,[18,19,20] or to an increased canting angle.[17] In our case, x-ray photoelectron spectroscopy results indicate that each cation, Bi, Fe and Cr composed the BFCO films were trivalent without the existence of divalent ferrous ions (Fig. 1 A), thus ruling out the divalent-based mechanism. Therefore, the increase in macroscopic magnetization with thickness could possibly be due to the increase of the canting angle and/or a re-ordering of the Cr and Fe cations.

4. Conclusion

In summary, we successfully prepared epitaxial BFCO films on Nb-doped STO (100), by pulsed laser deposition. The crystal structure found is very similar to that of BFO, and the films have the correct cationic stoichiometry throughout their thickness. The BFCO films exhibit good ferroelectric and piezoelectric properties at room temperature, a property that was not predicted by ab-initio calculations. Magnetic measurements show that the BFCO films exhibit a magnetic hysteresis at room temperature with a saturation magnetization about one order of magnitude higher than that of our BFO films having the same thickness.[7] Our results qualitatively confirm the predictions made using the *ab-initio* calculations about the existence of multiferroic properties in BFCO films. The existence of magnetic ordering at room temperature is an unexpected but very promising result that needs to be further investigated, and studies of the magnetic ordering and of the magnetoelectric coupling in the films are currently underway.

**ACKNOWLEDGMENTS**

The authors want to thank Dr. P. Plamondon ((CM)2, École Polytechnique de Montréal) for XTEM and SAED analysis and related discussions, to F. Normandin and Prof. T. Veres for performing preliminary magnetic measurements, and to thank Prof. R. W. Paynter for insightful comments about X-ray photoelectron spectroscopy. Part of this research was supported by INRS start-up funds, NRC-IMI operational funds, NSERC (Canada), and FQRNT (Québec).

**FIGURES CAPTIONS**

**Figure 1 A** XPS spectra at various depths from surface to interface of Fe2p and Cr2p core levels for a 80 nm thick BFCO film deposited on (100)-STO: Nb crystalline substrate as a function of the depth from the surface. The same region of the spectra of a 90 nm thick BFO layer ($Fe2p_{3/2}$ peak near 710.9 ± 0.2 eV) measured at a depth of 70 nm is also displayed for comparison.

**B** High-resolution XPS spectra of the O1s core level for BFO and BFCO films after a sputtering time of 30s: The experimental data is denoted by solid circles, the deconvoluted data by open symbols, and the fitted data by a solid line. The binding energy of the adsorbed oxygen is 532.8 eV and serves as a reference energy to accurately compute the energy shift of the films oxygen peak; the experimental error lies below ± 0.2 eV.

**Figure 2 A** X-ray $\theta/2\theta$ scan of the (00*l*) pseudocubic reflection for a BFCO layer deposited directly on (100)-oriented Niobium-doped STO substrate. **B** Φ-scan diffraction spectra of the (103) plane for the BFCO film and the STO substrate.

**Figure 3 A** Electron diffraction pattern of a 300nm-BFCO layer along the [011] direction.
**B** High-resolution TEM image of the epitaxial BFCO layer taken along [210] direction.

**Figure 4** Local piezoresponse switching of a single grain of epitaxial BFCO: **A** topography, **B** piezoresponse image before and **C** after switching of the grain. All of the images are 1 x 1 $\mu m^2$. **D** Local remanent and in field piezoresponse hysteresis obtained for epitaxial BFCO deposited on STO:Nb (100).

**Figure 5** Magnetic responses from BFCO films grown on 001 Nb doped STO with two different thicknesses.

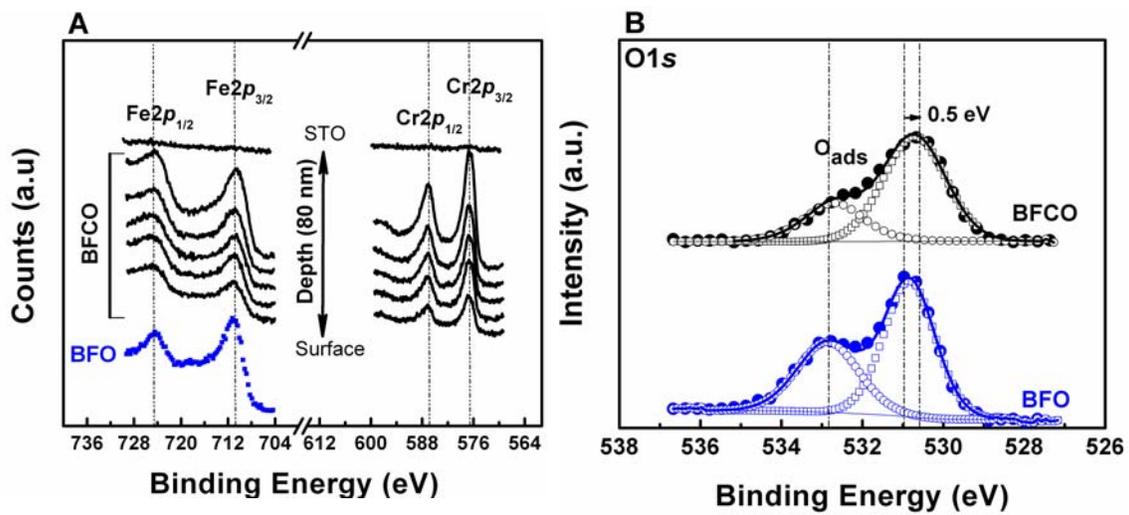

**Figure 1. R. Nechache et al.**

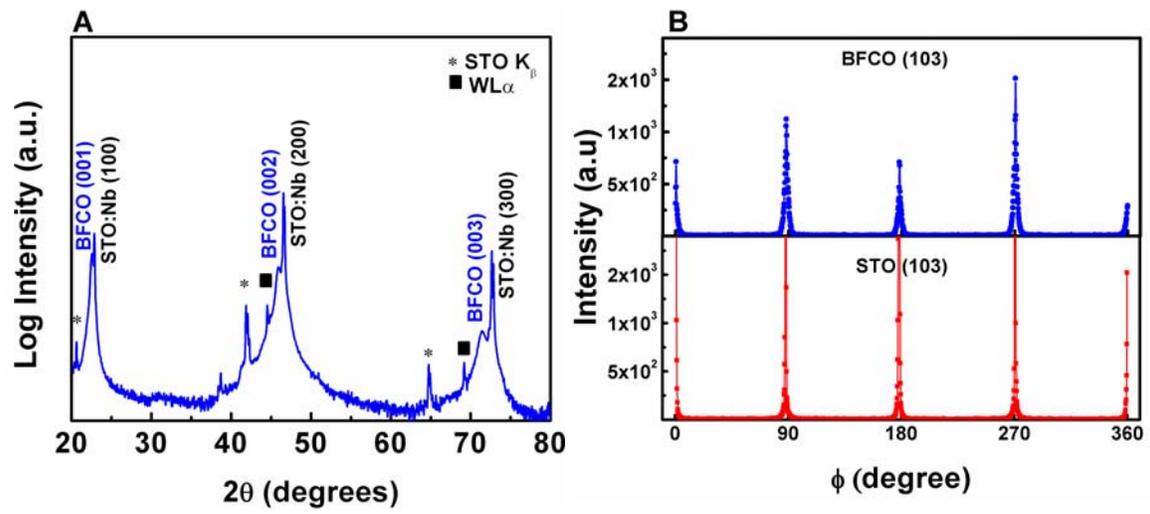

**Figure 2. R. Nechache et *al*.**

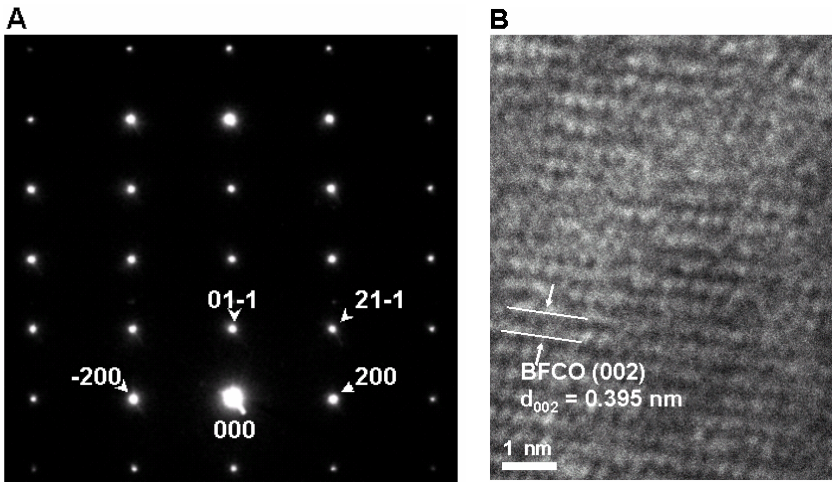

**Figure 3. R. Nechache et *al*.**

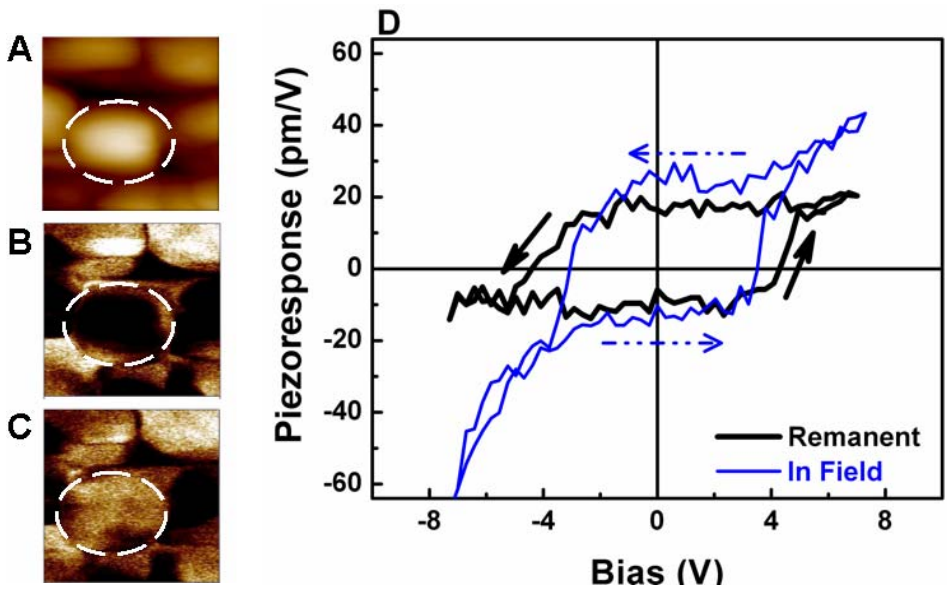

**Figure 4.** R. Nechache et *al*.

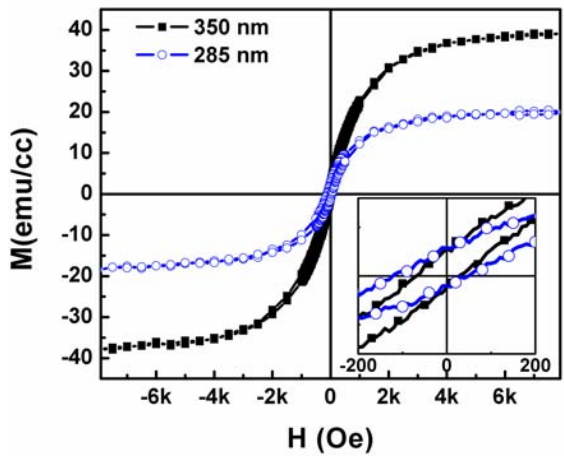

**Figure 5. R. Nechache et *al*.**